\documentclass[12pt]{article}
\usepackage{amsfonts}

\usepackage[colorlinks=true,citecolor=blue]{hyperref}

\begin{document}
	\title{The Fifth Force}
	\author{G. S. Burra\\
		Adj. Professor, University of Udine, 201, Via delle Scienze\\
		33100, Udine, Italy}
	\date{}
	\maketitle
	\Large
	\begin{abstract}
	Starting from dark energy considerations, we touch upon the Kerr-Newman 
	formulation, and this leads us to the hypothetical fifth force for which 
	hopeful confirmation is coming in.
	\end{abstract}
	\section{Introduction}
	 For  decades people have been thinking that there were only four 
	 fundamental forces in nature, but from over 20 years ago the author had 
	 been working with the idea  there could be a possible fifth short range 
	 force too. The author had published this in \cite{uof,cu} and also 
	 presented it in invited talks in Europe (for example in University of 
	 Udine) and USA (Vanderbilt University). He was motivated by the concept  
	 an all-pervading zero point 
	 energy, which was later christened `` Dark Energy'' and its presence 
	 confirmed by Adam Reiss, Saul Permutter \cite{perlnature} and others 
	 through 
	 observation of Type 1 supernovae. Many prominent scientists  noted this 
	 fact of Dark Energy. In the words of Tony Leggett ``It is of course clear 
	 that your equation predicts an exponential (inflation-type) expansion of 
	 the current universe, hence acceleration. And it would have been nice if 
	 the Nobel committee had mentioned this $\cdots$

	 Now with 
	 theoretical and practical 
	 work, it seems that nature also has a hitherto unknown new and a fifth 
	 force. The first hint 
	 that there could be a new strong  short range force came in the author's 
	 formulation of charged elementary particles as Quantum Mechanical Kerr 
	 Newman black holes in the late 1990s \cite{cu}. Off and on there were 
	 unsubstantiated claims  of the detection of such a force, for example from 
	 FermiLab. Already from around 2016, a Hungarian team \cite{nature-hunga} 
	 had 
	 claimed the discovery of a new force. This was whetted by the University 
	 of 
	 California, Irvine. But most physicists remain skeptical about this. On 
	 the other hand, very recently this has  been observed by a multi 
	 national team at the LHC b collider in Geneva \cite{LHCb}. They
	 were observing the decay of the beauty quark. This 
	 vindicates the work of the author, who, for many years has been 
	  publishing work on the subject.
	  \section{Theoretical justification for the existence of a fifth 
	  force}\begin{enumerate} 
	\item 
	{\textbf{The Kerr Newmann Quantum Mechanical Black Hole}}\\
We consider
	distances of the order of the Compton wavelengh. At this
	level Quantum Mechanical phenomena like zitterbewegung, \cite{dirac-pqm} 
	negative
	energy solutions and luminal velocities come in to play. Taking a route 
	through
	relativistic vortices, monopoles and classical considerations, we will be 
	lead
	to the model of leptons and quarks as what may be called ``Quantum
	Mechanical Kerr-Newman Black Holes" (QMKNBH), wherein features of Quantum
	Mechanics and General Relativity are inextricably inter-woven.
	 In natural units the Kerr Newmann metric is given by (cf. ref. 
	 \cite{MWT-73}).
	 \begin{eqnarray} \label{eq:1} \nonumber ds^2 &= &- \frac{\Delta}{\rho^2} 
	 [dt - a \sin^2 \theta d \phi]^2\\
	&	& \nonumber + \frac{\sin^2 \theta}{\rho^2} [(r^2 + a^2) d \phi - a 
	dt]^2\\
 &	&+ \frac{\rho^2}{\Delta} dr^2 + \rho^2 d \theta^2,\end{eqnarray}
	 where, $a$ is the Compton wavelength and,
	 $$\Delta = r^2 - 2mr + a^2 + m^2 + e^2, \rho^2 \equiv r^2 + a^2 \cos^2 
	 \theta$$
	 At $r = a$ and $\theta = \pi/2$, $\Delta = 2a^2$ as both $e$ and $m < < a,$
	 and $\rho^2 = a^2.$\\
	 It can be seen from  equation (\ref{eq:1}) that in addition to the usual 
	 long-range force, there are other forces of the order of atleast $1/r^3.$ 
	 These 
	 are shorter and stronger as  $r$ becomes small.
	 \item{\textbf{Short-lived vector Bosons}}\\
	 It is well known that the Dirac equation \cite{greiner,bd} is
	 \begin{equation}
	 	(\gamma^\mu p_\mu - m) \psi = 0\label{e1}
	 \end{equation}
	 Here $\gamma^\mu$ are $4 \times 4$ matrices obeying the Clifford
	 algebra and $\psi$ is a 4 component wave function (spinor). $\psi$
	 can be written as,
	 \begin{equation}
	 	\psi = \left(\begin{array}{ll} \phi \\
	 		\chi\end{array}\right)\label{e2}
	 \end{equation}
	 where $\phi$ and $\chi$ are 2-component spinors, $\phi$ being the
	 ``large" or positive energy component of $\psi$ and $\chi$ is the
	 ``small" or negative energy component which is such that
	 \begin{equation}
	 	\chi \sim (\frac{v}{c})^2 \phi\label{e3}
	 \end{equation}
	 It is also known that this picture gets reversed at high energies
	 where $v \to c$ (Cf.refs.\cite{greiner,bd}).\\
	 We observe that (\ref{e1}) can be written as: \cite{feshbach,uheb}
	 $$\imath \hbar (\partial \phi / \partial t) = c \tau \cdot (p -
	 e/cA) \chi + (mc^2+e\phi)\phi ,$$
	 \begin{equation}
	 	\imath \hbar (\partial \chi / \partial t) = c \tau \cdot (p - e/cA)
	 	\phi + (-mc^2+e\phi)\chi.\label{3.29}
	 \end{equation}
	 We can see from (\ref{3.29}) that (in the absence of electromagnetic
	 subsection),
	 \begin{equation}
	 	t \to -t, \quad \phi \to -\chi\label{e5}
	 \end{equation}
	 Let us now consider intervals near the Compton scale, where as we
	 know $v \to c$, and $\chi$ no longer is the ``small" component.\\
	 At the Compton scale we have the phenomenon of Zitterbewegung or
	 rapid unphysical oscillation. It has been pointed out that in this
	 case, \cite{uof} that time can be modelled by a double Weiner
	 process and can be described as follows
	 \begin{equation}
	 	\frac{d_+}{dt} x (t) = {\bf b_+} \, , \, \frac{d_-}{dt} x(t) = {\bf
	 		b_-}\label{2ex1}
	 \end{equation}
	 where for simplicity we consider in the one dimensional case. This
	 equation (\ref{2ex1}) expresses the fact that the right derivative
	 with respect to time is not necessarily equal to the left
	 derivative. It is well known that (\ref{2ex1}) leads to the Fokker
	 Planck equations \cite{tduniv,joos}
	 $$
	 \partial \rho / \partial t + div (\rho {\bf b_+}) = V \Delta \rho
	 ,$$
	 \begin{equation}
	 	\partial \rho / \partial t + div (\rho {\bf b_-}) = - U \Delta
	 	\rho\label{2ex2}
	 \end{equation}
	 defining
	 \begin{equation}
	 	V = \frac{{\bf b_+ + b_-}}{2} \quad ; U = \frac{{\bf b_+ - b_-}}{2}
	 	\label{2ex3}
	 \end{equation}
	 We get on addition and subtraction of the equations in (\ref{2ex2})
	 the equations
	 \begin{equation}
	 	\partial \rho / \partial t + div (\rho V) = 0\label{2ex4}
	 \end{equation}
	 \begin{equation}
	 	U = \nu \nabla ln\rho\label{2ex5}
	 \end{equation}
	 It must be mentioned that $V$ and $U$ are the statistical averages
	 of the respective velocities and their differences. We can then
	 introduce the definitions
	 \begin{equation}
	 	V = 2 \nu \nabla S\label{2ex6}
	 \end{equation}
	 \begin{equation}
	 	V - \imath U = -2 \imath \nu \nabla (l n \psi)\label{2ex7}
	 \end{equation}
	 We refer the reader to Smolin \cite{smolin} for further details. We
	 now observe that the complex velocity in (\ref{2ex7}) can be
	 described in terms of a positive or uni directional time $t$ only,
	 but with a complex coordinate
	 \begin{equation}
	 	x \to x + \imath x'\label{2De9d}
	 \end{equation}
	 To see this let us rewrite (\ref{2ex3}) as
	 \begin{equation}
	 	\frac{dX_r}{dt} = V, \quad \frac{dX_\imath}{dt} = U,\label{2De10d}
	 \end{equation}
	 where we have introduced a complex coordinate $X$ with real and
	 imaginary parts $X_r$ and $X_\imath$, while at the same time using
	 derivatives with respect
	 to time as in conventional theory.\\
	 From (\ref{2ex3}) and (\ref{2De10d}) it follows that
	 \begin{equation}
	 	W = \frac{d}{dt} (X_r - \imath X_\imath )\label{2De11d}
	 \end{equation}
	 This shows that we can use derivatives with respect to the usual
	 time derivative with the complex space coordinates (\ref{2De9d}) 
	 (Cf.ref.\cite{bgsfpl162003}.\\
	 Generalizing (\ref{2De9d}), to three dimensions, we end up with not
	 three but four dimensions,
	 $$(1, \imath) \to (I, \tau),$$
	 where $I$ is the unit $2 \times 2$ matrix and $\tau$s are the Pauli
	 matrices. We get the special relativistic Lorentz
	 invariant metric at the same time.\\
	 That is,\\
	 \begin{equation}
	 	x + \imath y \to Ix_1 + \imath x_2 + jx_3 + kx_4,\label{Aa}
	 \end{equation}
	 where $(\imath ,j,k)$ momentarily represent the Pauli
	 matrices; and, further,
	 \begin{equation}
	 	x^2_1 + x^2_2 + x^2_3 - x^2_4\label{B}
	 \end{equation}
	 is invariant, thus establishing a one to one correspondence between
	 (\ref{Aa}) and Minkowski 4 vectors as shown by (\ref{B}).\\
	 In this description we would have from (\ref{Aa}), returning to the
	 usual notation,
	 \begin{equation}
	 	[x^\imath \tau^\imath , x^j \tau^j] \propto \epsilon_{\imath jk}
	 	\tau^k \ne 0\label{y}
	 \end{equation}
	 (No summation over $\imath$ or $j$) Alternatively, absorbing the 
	 $x^\imath$ and
	 $\tau^\imath$ into each other, (\ref{y}) can be written as
	 \begin{equation}
	 	[x^\imath , x^j] = \beta \epsilon_{\imath jk} \tau^k\label{xa}
	 \end{equation}
	 Equation (\ref{y}) and (\ref{xa}) show that the coordinates no
	 longer follow a commutative geometry. It is quite remarkable that
	 the noncommutative geometry (\ref{y}) has been studied by the author
	 in some detail (Cf.\cite{tduniv}), though from the point of view of
	 Snyder's minimum fundamental length, which he introduced to overcome
	 divergence difficulties in Quantum Field Theory. Indeed we are
	 essentially in the same situation, because for our positive energy
	 description of the universe, there is the minimum Compton wavelength
	 cut off for a meaningful description as is well known
	 \cite{bgsextn,schweber,newtonwigner}. Following Feshbach and Villars
	 (loc.cit) we consider (\ref{e2}) to describe a particle
	 anti-particle pair.\\
	 Proceeding further we could invoke the $SU (2)$ and consider the
	 gauge transformation \cite{taylor}
	 \begin{equation}
	 	\psi (x) \to exp [\frac{1}{2} \imath g \tau \cdot \omega (x)] \psi
	 	(x).\label{4.2}
	 \end{equation}
	 This is known to lead to a gauge covariant derivative
	 \begin{equation}
	 	D_\lambda \equiv \partial_\lambda - \frac{1}{2} \imath g \tau \cdot
	 	\bar{W}_\lambda,\label{4.3a}
	 \end{equation}
	 We are thus lead to vector Bosons $\bar{W}_\lambda$ and an
	 interaction like the weak interaction, described by
	 \begin{equation}
	 	\bar{W}_\lambda \to \bar{W}_\lambda + \partial_\lambda \omega - g
	 	\omega \Lambda \bar{W}_\lambda.\label{4.4}
	 \end{equation}
	 However, we are this time dealing, not with iso spin, but between
	 positive and negative energy states as in (\ref{3.29}) that is
	 particles and antiparticles. Also we must bear in mind that this new
	 non-electroweak force between particles and anti particles would
	 be short lived as we are at the Compton scale \cite{report}.\\
	 These considerations are also valid for the Klein-Gordon equation
	 because of the two component formulation developed by Feshbach and
	 Villars \cite{feshbach,uheb}. There too, we get equations like
	 (\ref{3.29}) except that $\phi$ and $\chi$ are in this case scalar
	 function. We would like to re-emphasize that our usual description
	 in terms of positive energy solutions is valid above the Compton
	 scale (Cf.refs.\cite{greiner,bd}). To put it another way, equation
	 (\ref{e2}) describes a new spinor in a "superspin" space.\\
	 Thus we are lead to a new short lived interaction (as we are near
	 the Compton scale), mediated by vector Bosons $\bar{W}$.
	 \item
	Let us start with the equations,
	 \begin{equation}
	 	g_{\mu \nu} = \eta_{\mu \nu} + h_{\mu \nu}, h_{\mu \nu} = \int
	 	\frac{4T_{\mu \nu}(t-|\vec x - \vec x'|, \vec x')}{|\vec x - \vec
	 		x'|} d^3 x'\label{He1da}
	 \end{equation}
	 where as usual,
	 \begin{equation}
	 	T^{\mu \nu} = \rho u^u u^v\label{He2da}
	 \end{equation}
	 lead, on using (\ref{He2da}) in (\ref{He1da}), to the mass,
	 spin, gravitational potential and charge of 
	 an
	electron, if we work at the Compton 
	 scale (Cf.
	\cite{cu} for details). Let us now apply
	 the macro Gravitoelectric and 
	Gravitomagnetic
	 equations to the above case. Infact these equations are
	 (Cf.ref.\cite{mashhoon}).
	 \begin{equation}
	 	\nabla \cdot \vec E_g \approx -4\pi \rho, \nabla \times \vec E_g
	 	\approx - \partial \vec H_g/\partial t, etc.\label{He3da}
	 \end{equation}
	 \begin{equation}
	 	\vec E_g = - \nabla \phi - \partial \vec A/\partial t, \quad \vec
	 	H_g = \nabla \times \vec A\label{He4da}
	 \end{equation}
	 \begin{equation}
	 	\phi \approx - \frac{1}{2} (g_{00} + 1), \vec A_\imath \approx
	 	g_{0 \imath},\label{He5da}
	 \end{equation}
	 The subscripts $g$ in the equations (\ref{He3da}) and (\ref{He4da})
	 are to indicate that the fields $E$ and $H$ in the
	 macro case do not really represent the 
	electromagnetic
	 field, but rather resemble them. Let us apply equation
	 (\ref{He4da}) to equation (\ref{He1da}), keeping in mind equation
	 (\ref{He5da}). We then get, considering only the order of
	 magnitude, which is what interests us here, after some
	 manipulation
	 \begin{equation}
	 	|\vec H | \approx \int \frac{\rho V}{r^2} \bar r \approx \frac{m
	 		V}{r^2}\label{He6da}
	 \end{equation}
	 and
	 \begin{equation}
	 	| \vec E | = \frac{mV^2}{r^2}\label{He7da}
	 \end{equation}
	 $V$ being the speed.\\
	 In (\ref{He6da}) and (\ref{He7da}) the distance $r$ is much greater
	 than a typical Compton wavelength, to make the
	 approximations considered in
	 deriving the Gravitomagnetic and 
	 Gravitoelectric equations meaningful.\\
	 Remembering that we have, by the Uncertainty 
	 Principle,
	 $$mVr \approx h,$$
	 the electric and magnetic fields in (\ref{He6da}) and (\ref{He7da})
	 now become
	 \begin{equation}
	 	|\vec H | \sim \frac{h}{r^3} , |\vec E | \sim
	 	\frac{hV}{r^3}\label{He8da}
	 \end{equation}
	 We now observe that (\ref{He8da}) does not really contain the
	mass of the
	 elementary particle. Could we get a further insight into this new force?\\
	 Indeed in \cite{taylor}
	 characterization of the electron, it turns out as
	 indicated that theelectron can be represented by the
	Kerr-Newman metric which incidentally also gives the
	 anomalous gyromagnetic ratio $g=2$. (This result has more  recently been
	 reconfirmed by Nottale \cite{nottalecsf01} from a totally
	 different point of view, using scaled relativity). It is well
	 known that the Kerr-Newman field has extra electric and
	 magnetic terms (Cf.\cite{rr76}),
	 both of the order $\frac{1}{r^3}$, exactly as indicated in (\ref{He8da}).\\
	 It may be asked if there is any candidate as yet for the above
	 mass independent, spin dependent (through $h$)
	 short range force. Perhaps the speculated inexplicable $B_{(3)}$ 
	 \cite{ar14e} short range
	 force could be a candidate. It differs from the
	 usual $B_{(1)}$ and $B_{(2)}$ long range fields of Special Relativity.\\
	 Interestingly, if we think of the above force as being mediated by
	 a ``massive'' particle, that is, work with a massive
	 vector field we can recover (\ref{He7da}) and
	 (\ref{He8da}) \cite{iz}. 
	 A final comment: It is quite remarkable that equations like
	 (\ref{He3da}), (\ref{He4da}) and (\ref{He5da}) which resemble the
	 equations of electromagnetism, have in the usual 
	 macro
	 considerations no connection whatsoever with
	electromagnetism except in appearance. This would 
	 seem to
	 be a rather miraculous coincidence. In fact the above
	 considerations of  the 
	 Kerr-Newman
	 metric formulation, demonstrate that the resemblance to
	 electromagnetism is not an accident, because in 
	 this
	 latter formulation, both electromagnetism and
	gravitation arise from the metric (Cf.also
	 refs.\cite{ar4,annales,ar8e,cu}).
	 \item We can arrive at the same conclusion from a slightly different point 
	 of view.
	 
	 We recall  the equations 
	 \begin{equation}
	 	g_{\mu \nu} = \eta_{\mu \nu} + h_{\mu \nu}, h_{\mu \nu} = \int
	 	\frac{4T_{\mu \nu}(t-|\vec x - \vec x'|, \vec x')}{|\vec x - \vec
	 		x'|} d^3 x'\label{8He1da}
	 \end{equation}
	 where as usual,
	 \begin{equation}
	 	T^{\mu \nu} = \rho u^u u^v\label{8He2da}
	 \end{equation}
	 lead, on using (\ref{8He2da}) in (\ref{8He1da}), to the
	 mass, spin, gravitational
	 potential and charge of an electron, if we work at
	 the Compton scale. Let us now apply the macro
	 Gravitoelectric and
	 Gravitomagnetic equations to the above case.
	 Infact these equations are (Cf.ref.\cite{mashhoon}).
	 \begin{equation}
	 	\nabla \cdot \vec E_g \approx -4\pi \rho, \nabla \times \vec E_g
	 	\approx - \partial \vec H_g/\partial t, etc.\label{8He3da}
	 \end{equation}
	 \begin{equation}
	 	\vec E_g = - \nabla \phi - \partial \vec A/\partial t, \quad \vec
	 	H_g = \nabla \times \vec A\label{8He4da}
	 \end{equation}
	 \begin{equation}
	 	\phi \approx - \frac{1}{2} (g_{00} + 1), \vec A_\imath \approx g_{0
	 		\imath},\label{8He5da}
	 \end{equation}
	 The subscripts $g$ in the equations (\ref{8He3da}) and
	 (\ref{8He4da})
	 are to indicate that the fields $E$ and $H$ in the
	 macro case do not really represent the
	electromagnetic field, but rather resemble
	 them. Let us apply equation (\ref{8He4da}) to equation
	 (\ref{8He1da}), keeping in mind equation (\ref{8He5da}). We then
	 get, considering only the order of magnitude, which is what
	 interests us here, after some manipulation
	 \begin{equation}
	 	|\vec H | \approx \int \frac{\rho V}{r^2} \bar r \approx \frac{m
	 		V}{r^2}\label{8He6da}
	 \end{equation}
	 and
	 \begin{equation}
	 	| \vec E | = \frac{mV^2}{r^2}\label{8He7da}
	 \end{equation}
	 $V$ being the speed.\\
	 In (\ref{8He6da}) and (\ref{8He7da}) the distance $r$ is much
	 greater than a typical Compton wavelength,
	 to make the approximations considered in deriving the
	 Gravitomagnetic and
	 Gravitoelectric equations meaningful.\\
	 Remembering that we have, by the Uncertainty Principle,
	 $$mVr \approx h,$$
	 the electric and magnetic fields in (\ref{8He6da}) and
	 (\ref{8He7da}) now become
	 \begin{equation}
	 	|\vec H | \sim \frac{h}{r^3} , |\vec E | \sim
	 	\frac{hV}{r^3}\label{8He8da}
	 \end{equation}
	 We now observe that (\ref{8He8da}) does not really contain the
	 mass of the
	 elementary particle. Could we get a further insight into this new force?\\
	 Indeed in the above linearized General Relativistic characterization
	 of the electron, it turns out as indicated that the
	 electron can be represented by the
	Kerr-Newman metric which incidentally also
	 gives the anomalous gyromagnetic ratio $g=2$ of ref. \cite{cu}.
	 (This result has recently been reconfirmed by Nottale
	 \cite{nottalecsf01} from a totally different point of view, using
	 scaled relativity). It is well known that the
	 Kerr-Newman field has extra electric and magnetic
	 terms (Cf.\cite{rr76}),
	 both of the order $\frac{1}{r^3}$, exactly as indicated in 
	 (\ref{8He8da}).\\ which we got beforeIncidentally this result can 
	  be related to the care new man metric whatever it is, We are lead
	 Two a short strong force which is at least off the order of $1/r^3$
  \end{enumerate}
	  \section{Experimental evidence} 
	  \subsection\\
	  In recent years, a group of Hungarian researchers claimed that  they 
	  discovered a new particle — called X17. Such a particle would require the 
	  existence of a fifth force of nature.
	Some physicists are sceptical that the new particle exists. 
	  
	  Confirmation for  such a 17-MeV particle is awaited. 
	 The Jefferson Laboratory  in the DarkLight experiment aims to search for 
	 dark 
	  photons with masses of 10–100 MeV (cf. also \cite{bgs-mevdarkmatter}). 
	  They 
	  propose to do this by  firing 
	  electrons at a hydrogen gas target.
	  They hope that they would be able 
	  to either find the proposed particle or find suitable limits on its 
	  coupling with normal matter.
	  They have now seen it show up in the 
	  same way hundreds of times.
	  That leaves some physicists excited by the prospect of a new force. But 
	  even if 
	  an unknown force is not responsible for the strange signal, the team may 
	  have 
	  revealed some novel, hitherto and unknown physics even help explain dark 
	  matter. 
	  However, so far, most scientists remain skeptical. For years, researchers 
	  tied 
	  to the Hungarian group have claimed to discover new particles that later 
	  vanished.  
	  The group shot protons at 
	  a thin sample of Lithium-7, which then 
	  radioactively decayed into Beryllium-8. As expected, this created pairs 
	  of 
	  positrons and electrons. However, the detectors also picked up excess 
	  decay 
	  signals that suggested the existence of a potential new and extremely 
	  weak 
	  particle. If it exists, the particle would be about 1/50 the 
	  mass of a 
	  proton. And because of its properties, it would be a boson. The findings 
	  were peer viewed and published in the Physical 
	  review letters \cite{kras, nature-hunga,feng}.
	  
	 \subsection{}
	  \cite{LHCb}
	  The Large Hadron Collider (LHC) physicists, in March 2021, reported  
	  signals 
	  for new physics - a new force of nature. We come to Beauty quarks, or 
	  bottom quarks. A paper put out by the 
	  collaboration im March  was based on data from the LHCb experiment, 
	  that recorded the outcome of the ultra high-energy collisions produced by 
	  the LHC and found that beauty quarks were decaying into electrons and  
	  muons 
	  at different rates, which should not be so. When a beauty quark decays 
	  into electrons or muons via the weak force, it ought to do so equally 
	  often. Instead, it was found that the muon decay was only happening about 
	  85\% as often as the electron decay. This would imply some new  force of 
	  nature that pulls on electrons and 
	  muons differently is interfering with how beauty quarks decay. The result 
	  was a``three $\sigma$” one.. The  study 
	  so far examined  beauty quarks that were paired up with ``up” quarks.  
	  Two other decays were also studied: one where the beauty quarks that were 
	  paired 
	  with ``down” quarks and another where they were also paired with ``up" 
	  quarks. This time, muon decays were only happening around 70\% as often 
	  as the electron decays. This time the result was a two $\sigma$ result. 
	  We might be on the brink of a major breakthrough, but more data is 
	  required. In addition, other experiments at the LHC, as well at the Belle 
	  2 experiment in Japan, are also looking for these results.


\begin{thebibliography}{99}
	\bibitem{uof} Sidharth, B. G. (2005)  \emph{The universe of 
		fluctuations}, Springer Netherlands.
		\bibitem {cu} Sidharth, B.G. (2001), \emph{Chaotic Universe: From the 
		Planck to the Hubble Scale}
	(Nova Science, New York).
		\bibitem {perlnature} Perlmutter, S.,  et al. (1998). \emph{Nature} 
	Vol.391, 1 January 1998, pp.51--59.
		\bibitem{nature-hunga}Cartlidge, E. (2016) \emph{``Has a Hungarian 
		physics lab found 
	a fifth force of nature?."}  Nature News.
	\bibitem{LHCb} LHCb collaboration, Aaij R., Abdelmotteleb  A. S. W.  
	et. al., https://arxiv.org/abs/2110.09501.
	\bibitem{dirac-pqm}Dirac, P.A.M., 1981. \emph{The principles of quantum 
	mechanics} (No. 27). Oxford university press.
\bibitem {MWT-73} Misner,C.W. ,  Thorne  K.S. and Wheeler  J.A., (1973)
\emph{``Gravitation"}, W.H. Freeman, San Francisco, , pp.819ff.
\bibitem {greiner} Greiner, W. and Reinhardt, J. (1987). 
\emph{Relativistic Quantum Mechanics:
	Wave Equation} 2nd Ed. (Springer, New York,1987).
	\bibitem {bd} Bjorken, J.D. and Drell, S.D. (1965). \emph{Relativistic 
	Quantum Fields}
(McGraw-Hill Inc., New York), pp.44ff.
\bibitem {feshbach} Freshbach, H. and Villars, F. (1958).
Rev.Mod.Phys. Vol.30, No.1, January 1958, pp.24-45.
\bibitem {uheb} Sidharth, B.G. (2011) \emph{Ultra High Energy
	Behaviour}, arXiv:1103.1496.
	\bibitem {tduniv} Sidharth, B.G. (2008). \emph{The Thermodynamic 
	Universe} (World Scientific, Singapore).
\bibitem {joos} Joos, G. (1951). \emph{Theoretical Physics} (Blackie, 
London), pp199ff.
	\bibitem {smolin} Smolin, L. (1986). \emph{Quantum Concepts in Space 
	and Time},
Penrose, R. and  Isham, C.J. (eds.) (OUP, Oxford), pp.147--181.
	\bibitem {bgsfpl162003} Sidharth, B.G. (2003) \emph{Found.Phys.Lett.} 
16, (1), pp.91--97.
	\bibitem {bgsextn} Sidharth, B.G. (2002). \emph{A brief note on 
	``extension, spin and non-commutativity”} Foundation of Physics 
	Letters 15 (5), 2002, 501ff.
	\bibitem {schweber} S.S. Schweber. (1961). \emph{An Introduction to 
	Relativistic Quantum
	Field Theory} (Harper and Row, New York).
\bibitem {newtonwigner} Newton, T.D. and Wigner, E.P. (1949) \emph{Localized 
States for Elementary Systems},
Reviews of Modern Physics, Vol.21,
No.3, July 1949, pp.400-405.
\bibitem {taylor} Taylor, J.C. (1976). \emph{Gauge Theories of Weak
	Interactions} (Cambridge University Press, Cambridge) 1976.
\bibitem {report} Sidharth, B.G. (2011). \emph{Brief Report} in
\emph{New Advances in Physics} Vol.5 (1), 2011, pp.49-50.
	\bibitem {mashhoon} Mashhoon, B., Ho Jung Paik. and Clifford M. Will. 
(1989). Physical Review D
	\emph{Detection of the gravitomagnetic field using an orbiting 
	superconducting gravity gradiometer. Theoretical principles.} 39.10 : 2825.

\bibitem{nottalecsf01} Nottale, L., (2001) Chaos, Solitons and Fractals, 
{\bf 12}  (9),pp.1577ff.
	\bibitem {rr76} Newman, E.T., J.Math.Phys.\underline{14} (1), 1973, p102

	\bibitem {ar14e}  Evans, M.W., ( 1997), ``Origin, Observation and 
	Consequences 
of the $B^{(3)}$ Field" in The Present Status of the Quantum Theory of
Light, S. Jeffers et al. (eds), Kluwer Academic Publishers,
Netherlands, pp.117-125 and several other references
therein.
	\bibitem {iz} Itzkson, C., Zuber, J., 1980, ``Quantum Field Theory", McGraw
Hill, New York,  pp.134-138.

	\bibitem {ar4} Sidharth, B.G., (2001) \emph{Towards a Unified Description 
	of the Fundamental Interactions} Chaotic Universe,Nova Science,New York.
	https://arxiv.org/abs/quant-ph/9805013.
		\bibitem {annales} Sidharth, B.G., (2002)  Annales de la Fondation Lois 
		de 
	Broglie, Volume 27 No.2,  333-342.
	\bibitem {ar8e} Sidharth, B.G., 2001, Nuovo Cimento, {\bf 116B} (6),  
	p.4ff.
	\bibitem{bgs-mevdarkmatter}Sidharth, B. G. (2005)\emph{A brief note on MeV 
	dark matter} International Journal of Modern Physics E, 14(04), 631-633.

	\bibitem{kras}  Krasznahorkay, A. J. et al. (2016) \emph{Observation of 
	Anomalous Internal Pair Creation in 8Be: A Possible Indication of a Light, 
	Neutral Boson} Physical Review 	Letters 116, 042501 .
\bibitem{feng} Feng, J. L. et al. (2016) \emph{Protophobic Fifth-Force 
	Interpretation 
	of the Observed Anomaly in $^{8}\mathrm{Be}$ Nuclear Transitions}, 
Phys. Rev. 
Lett., 117 (7),071803.
	


	
	\bibitem {nottalecsf94} Nottale, L. (1994). \emph{Chaos, Solitons \& 
		Fractals} 4, 3, pp.361--388
	and references therein.
	\bibitem {ness} Sidharth, B.G. (2011). \emph{Negative Energy
		Solutions and Symmetries}, \emph{arXiv:1104.0116}.
	

	
	
	\bibitem {bgsijmpe} Sidharth, B.G. (2005). \emph{Int.J.Mod.Phys.E.}
	14 (6), 2005, pp.923ff.

\end{thebibliography}
\end{document}